\documentclass[epj]{svjour}
\usepackage{graphics} 
\usepackage{graphicx} 
\usepackage{epstopdf}
\usepackage{color}
\usepackage{scrtime}
\usepackage{amssymb}
\usepackage{amsmath}
\usepackage{amsfonts}
\usepackage[T1]{fontenc}
\usepackage{dcolumn}
\usepackage{bm}
\usepackage[ulem=normalem]{changes}
\usepackage{xfrac}
\begin{document}

\title{Neutron--proton spin--spin correlations\\
in the ground states of $N=Z$ nuclei}
\author{P.~Van~Isacker\inst{1} \and A.~O.~Macchiavelli\inst{2}} 
%
%%\offprints{}          % Insert a name or remove this line
%
\institute{Grand Acc\'el\'erateur National d'Ions Lourds, CEA/DRF--CNRS/IN2P3, Bvd Henri Becquerel, F-14076 Caen, France
\and
Nuclear Science Division, Lawrence Berkeley National Laboratory, Berkeley, California 94720, USA}
\date{\today}
\abstract{
We present expressions for the matrix elements
of the spin--spin operator $\vec S_{\rm n}\cdot\vec S_{\rm p}$
in a variety of coupling schemes.
These results are then applied to calculate
the expectation value $\langle\vec S_{\rm n}\cdot\vec S_{\rm p}\rangle$
in eigenstates of a schematic Hamiltonian
describing neutrons and protons interacting in a single-$l$ shell through a Surface Delta Interaction.
The model allows us to trace $\langle\vec S_{\rm n}\cdot\vec S_{\rm p}\rangle$
as a function of the competition between the isovector and isoscalar interaction strengths
and the spin--orbit splitting of the $j=l\pm{\sfrac12}$ shells.
We find negative $\langle\vec S_{\rm n}\cdot\vec S_{\rm p}\rangle$ values
in the ground state of all even--even $N=Z$ nuclei,
contrary to what has been observed in hadronic inelastic scattering at medium energies.
We discuss the possible origin of this discrepancy
and indicate directions for future theoretical and experimental studies
related to neutron--proton spin--spin correlations.
\PACS{
      {PACS-key}{discribing text of that key}   \and
      {PACS-key}{discribing text of that key}
     } % end of PACS codes
} %end of abstract
\maketitle
\section{Introduction}
The nuclear pairing mechanism~\cite{Bohr58} has been, for many years,
a central subject of study in low-energy nuclear physics~\cite{Broglia13}. 
Although the energy gain of the nuclear system due to pairing is relatively modest,
pairing correlations have a strong influence on many properties of the nucleus
including the moment of inertia, deformation and excitation spectra~\cite{Brink}.
The dominant pairing in almost all known nuclei with $N>Z$ is
that in which "superconducting" pairs of neutrons (nn) and protons (pp)
couple to a state with angular momentum zero and isospin $T=1$,
known as isovector or spin-singlet pairing.  
However, for nuclei with $N\approx Z$,
neutrons and protons occupy the same single-particle orbits
at their respective Fermi surfaces
and Cooper pairs, consisting of a neutron and a proton (np), may form.
These types of pairs may couple
in either isovector or isoscalar (spin-triplet with  $J=1$ and $T=0$) modes,
the latter being allowed by the Pauli principle.
Contrary to the case of nuclei with large isospin imbalance,
where the spin--orbit suppresses pairing in the triplet channel,
in nuclei with $N=Z$ the isoscalar mode is expected to dominate.
Since the nuclear force is charge independent,
one would also expect that pairing should manifest
equivalently for np pairs with $T=1$ and $S=0$, akin to nn and pp pairs.
While there are convincing arguments for the existence of isovector np pairs,
the existence of a correlated isoscalar np  pair in condensate form,
and the magnitude of such collective pairing,
remains an intriguing and controversial topic in nuclear-structure physics~\cite{Frauendorf14}.

Long-standing theoretical predictions of the onset of isoscalar pairing,
the interplay between both pairing modes,
and the presence of a condensate composed of both isoscalar and isovector pairs
have remained without experimental confirmation~\cite{Frauendorf14,aom,Afanasejev13}.
This is mainly because the region of the nuclear landscape near the proton drip line,
where such phenomena are expected to appear, is largely unreachable
and because the experimental observables
are either inconclusive and/or complicated to interpret.
Two-neutron transfer reactions such as (p,t) and (t,p)
have provided a key probe to understand neutron pairing correlations in nuclei~\cite{Broglia73,Bayman68}.
The rapid quenching of np pairs as one moves away from $N=Z$~\cite{Engel96}
suggests that the transfer of a np pair from even--even to odd--odd self-conjugate nuclei
could also be a sensitive tool to study np correlations.
Hence, reactions such as ($^3$He,p) and (p,$^3$He) are among the best choices~\cite{Frobrich71,Isacker05}. 
\begin{figure}
\centering
\includegraphics[width=14cm,angle=0]{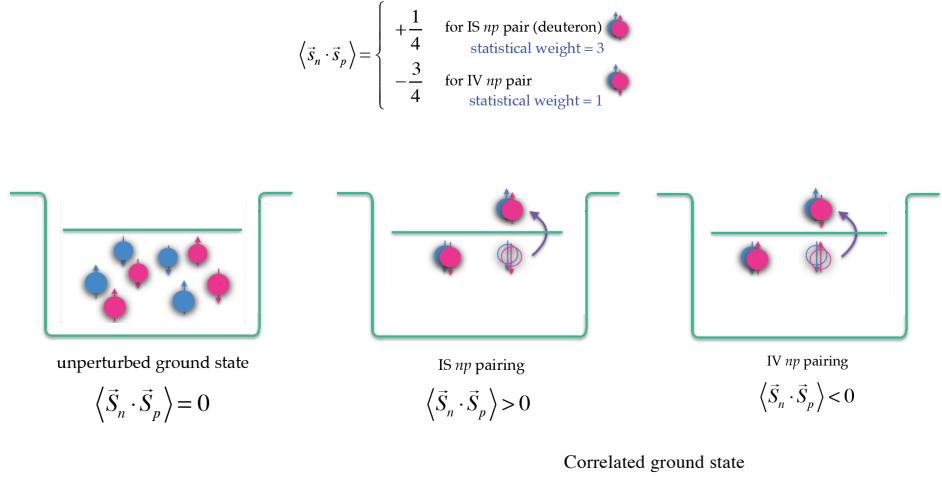}
\caption{
Schematic illustration (in $LS$ coupling)
of the dependence of the observable $\langle \vec S_{\rm n}\cdot \vec S_{\rm p}\rangle $
on the isoscalar (IS) or isovector (IV) nature of pairing correlations in the ground state.
Adapted from Ref.~\cite{Tamii15}.}
\label{fig1}
\end{figure}

A different and elegant approach has been proposed by the Osaka group~\cite{Matsubara10, Matsubara15}
and consists of the study of neutron--proton spin--spin correlations
in the ground states of $N\approx Z$ nuclei. 
The relevant observable is $\langle\vec S_{\rm n}\cdot\vec S_{\rm p}\rangle$,
the scalar product between the total spins of the neutrons and protons,
which can be measured by spin $M1$ excitations
produced by inelastic hadronic scattering at medium energies.
In Fig.~\ref{fig1} we illustrate why this quantity can inform us
on the nature of the pairing condensate.
It can be seen that given the distinctive values in the two-particle system,
$\langle\vec S_{\rm n}\cdot\vec S_{\rm p}\rangle $ will also depend strongly
on the type of pairs being scattered across the Fermi surface,
as will be discussed in Sect.~\ref{s_mes}. 

In a series of experiments carried out at the RCNP~\cite{Matsubara15} facility,
high-energy-resolution proton inelastic scattering $E_{\rm p}=295$~MeV
was studied in $^{24}$Mg, $^{28}$Si, $^{32}$S and $^{36}$Ar.
The results give positive values of
$\langle\vec S_{\rm n}\cdot\vec S_{\rm p}\rangle\approx0.1$ for the $sd$ shell
suggesting a predominance of {\em quasi-deuterons},
somewhat at variance with the discussions above
and USD shell-model calculations that are unable to reproduce the experimental results.
However, shell-model wave functions
that take into account an enhanced spin-triplet pairing
seem to reproduce the measured spin--spin correlations~\cite{Sagawa18}.
Also, the no-core shell model with realistic interactions~\cite{Barrett13}
predicts positive values (lower limits due to convergence)
that could be attributed to mixing with higher-lying orbits due to the tensor correlation.   

It seems clear to us that further work is required
to fully assess the origin of the spin--spin correlation and its microscopic origin. 
For example:
Are the observed spin--spin correlations between neutrons and protons connected to
a) our beloved surface pairing BCS condensate~\cite{Bohr58},
b) aligned np pairs~\cite{Cederwall11}
or c) effects of the tensor force~\cite{Subedi08}? are questions that remain to be answered.

To shed light on these questions,
we develop in this work the formalism to calculate the matrix elements
of the $\vec S_{\rm n}\cdot\vec S_{\rm p}$ operator in a variety of coupling schemes
and apply it to the solution of a schematic model
consisting of nucleons in a single-$l$ shell.
In spite of its simplicity, the model allows us
to study the behaviour of $\langle\vec S_{\rm n}\cdot\vec S_{\rm p}\rangle$
as a function of the competition
between the isovector and isoscalar components of the effective force between nucleons,
and the spin--orbit splitting of the $j=l\pm{\sfrac12}$ shells.
In Sect.~\ref{s_operator} we discuss the structure of the $\vec S_{\rm n}\cdot\vec S_{\rm p}$ operator
and we calculate its matrix elements in Sect.~\ref{s_mes}. 
In Sect.~\ref{s_sm}, following a short discussion of the model, 
we present and discuss our results for several cases
involving particles occupying shells with $l=1$ to 5
and contrast these with the experimental observations to date.
Finally, Sect.~\ref{s_sum} is devoted to the summary and conclusions of our work.

\section{The $\vec S_{\rm n}\cdot\vec S_{\rm p}$ operator}
\label{s_operator}
The $\vec S_{\rm n}\cdot\vec S_{\rm p}$ operator is given by
\begin{equation}
\vec S_{\rm n}\cdot\vec S_{\rm p}=
\sum_{k\in\{{\rm n}\}}\sum_{k'\in\{{\rm p}\}}\vec s(k)\cdot\vec s(k'),
\label{e_ss1}
\end{equation}
where the sums are over the neutrons and over the protons in the nucleus.
Introducing the isospin projection operator $t_z$,
which gives $+\sfrac12$ acting on a neutron and $-\sfrac12$ acting on a proton,
we rewrite this operator as
\begin{equation}
\vec S_{\rm n}\cdot\vec S_{\rm p}=
\sum_{kk'}\left({\textstyle{\frac12}}+t_z(k)\right)\left({\textstyle{\frac12}}-t_z(k')\right)\vec s(k)\cdot\vec s(k')=
{\frac14}\sum_{kk'}\vec s(k)\cdot\vec s(k')-
\sum_{kk'}t_z(k)t_z(k')\vec s(k)\cdot\vec s(k'),
\label{e_ss2}
\end{equation}
where the sums are over all nucleons in the nucleus.
It follows that $\vec S_{\rm n}\cdot\vec S_{\rm p}$
contains an isoscalar as well as an isotensor part.

Let us consider the case of nucleons occupying a single-$l$ shell.
We introduce the spin, isospin and spin--isospin operators
\begin{align}
T^{(010)}_{0\mu0}={}&
\sqrt{2(2l+1)}(a_{l{\sfrac12}{\sfrac12}}^\dag\times\tilde a_{l{\sfrac12}{\sfrac12}})^{(010)}_{0\mu0},
\nonumber\\
T^{(001)}_{00\nu}={}&
\sqrt{2(2l+1)}(a_{l{\sfrac12}{\sfrac12}}^\dag\times\tilde a_{l{\sfrac12}{\sfrac12}})^{(001)}_{00\nu},
\nonumber\\
T^{(011)}_{0\mu\nu}={}&
\sqrt{2l+1}(a_{l{\sfrac12}{\sfrac12}}^\dag\times\tilde a_{l{\sfrac12}{\sfrac12}})^{(011)}_{0\mu\nu},
\label{e_su4}
\end{align}
in terms of the nucleon creation operators $a_{lm_l{\sfrac12}m_s{\sfrac12}m_t}^\dag$
and the modified annihilation operators
$\tilde a_{lm_l{\sfrac12}m_s{\sfrac12}m_t}\equiv-(-)^{l+m_l+m_s+m_t}a_{l-m_l{\sfrac12}-m_s{\sfrac12}-m_t}$.
The operators~(\ref{e_su4}) are scalar with respect to the orbital angular momentum
and generate Wigner's SU(4) supermultiplet algebra~\cite{Wigner37}.
The representation~(\ref{e_ss2}) shows that $\vec S_{\rm n}\cdot\vec S_{\rm p}$
can be written as
\begin{equation}
\vec S_{\rm n}\cdot\vec S_{\rm p}=
{\frac14}\sum_\mu T^{(010)}_{0\mu0}T^{(010)}_{0\mu0}-
\sum_\mu T^{(011)}_{0\mu0}T^{(011)}_{0\mu0},
\label{e_sssq}
\end{equation}
which proves that it is an element of the SU(4) algebra.
The SU(4) tensor character of $\vec S_{\rm n}\cdot\vec S_{\rm p}$ is derived in the Appendix.

\section{Matrix elements of the $\vec S_{\rm n}\cdot\vec S_{\rm p}$ operator}
\label{s_mes}
One-body matrix elements of $\vec S_{\rm n}\cdot\vec S_{\rm p}$ vanish,
\begin{equation}
\langle lm_lsm_stm_t|\vec S_{\rm n}\cdot\vec S_{\rm p}|lm_lsm_stm_t\rangle=
\langle lsjm_jtm_t|\vec S_{\rm n}\cdot\vec S_{\rm p}|lsjm_jtm_t\rangle=0.
\label{e_me1}
\end{equation}

Two-body matrix elements can be derived in $LS$ or in $jj$ coupling,
and in both cases in an isospin or in a neutron--proton basis.
Since $\vec S_{\rm n}\cdot\vec S_{\rm p}$
is a scalar in orbital angular momentum, spin and total angular momentum,
the associated projections $M_L$, $M_S$ and $M_J$ can be suppressed.
It is, however, not a scalar in isospin
and therefore its matrix elements depend on the projection $M_T$.
In an $LST$ basis the two-body matrix elements are
\begin{align}
\langle l^2LSTM_T=0|\vec S_{\rm n}\cdot\vec S_{\rm p}|l^2LSTM_T=0\rangle={}&
{\frac14}[2S(S+1)-3],
\nonumber\\
\langle l^2LSTM_T=\pm1|\vec S_{\rm n}\cdot\vec S_{\rm p}|l^2LSTM_T=\pm1\rangle={}&0,
\label{e_me2lst}
\end{align}
where it is assumed that $L+S+T$ is odd.
In a $JT$ basis the two-body matrix elements are
\begin{align}
\langle j^2JTM_T=0|\vec S_{\rm n}\cdot\vec S_{\rm p}|j^2JTM_T=0\rangle={}&
{\frac12}\sum_{LS}S(S+1)
\left[\begin{array}{ccc}
l&{\sfrac12}&j\\l&{\sfrac12}&j\\L&S&J
\end{array}\right]-{\frac34},
\nonumber\\
\langle j^2JTM_T=\pm1|\vec S_{\rm n}\cdot\vec S_{\rm p}|j^2JTM_T=\pm1\rangle={}&0,
\label{e_me2jt}
\end{align}
where it is assumed that $J+T$ is odd
and that the sum runs over odd $L+S+T$.

Equation~(\ref{e_me2jt}) can be applied if the two nucleons are in the same $j$ shell.
If the nucleons occupy an $l$ shell,
matrix elements of $\vec S_{\rm n}\cdot\vec S_{\rm p}$ are needed
with one nucleon in the $l+{\sfrac12}$ and the other in the $l-{\sfrac12}$ shell.
In this case it is more convenient to consider the problem in a neutron--proton basis.
The expression for the two-body matrix elements of $\vec S_{\rm n}\cdot\vec S_{\rm p}$
is particularly simple in an $LS$-coupled neutron--proton basis,
where the only non-zero matrix element is
\begin{equation}
\langle(l_{\rm n}{\sfrac12})(l_{\rm p}{\sfrac12})LS|
\vec S_{\rm n}\cdot\vec S_{\rm p}
|(l_{\rm n}{\sfrac12})(l_{\rm p}{\sfrac12})LS\rangle=
(-)^{S+1}{\frac32}
\biggl\{\begin{array}{ccc}
{\sfrac12}&{\sfrac12}&1\\{\sfrac12}&{\sfrac12}&S
\end{array}\biggr\}=
\Biggl\{
\begin{array}{l}
-{\displaystyle\frac34},\quad S=0,\\[5pt]
+{\displaystyle\frac14},\quad S=1.\end{array}
\label{e_me2lsnp}
\end{equation}
In a $jj$-coupled neutron--proton basis
the matrix element of $\vec S_{\rm n}\cdot\vec S_{\rm p}$ is
\begin{equation}
\langle j_{\rm n} j_{\rm p} J|\vec S_{\rm n}\cdot\vec S_{\rm p}|j'_{\rm n} j'_{\rm p} J\rangle=
(-)^{j'_{\rm n}+j_{\rm p}+J}
\langle j_{\rm n}\|\vec S_{\rm n}\|j'_{\rm n}\rangle
\langle j_{\rm p}\|\vec S_{\rm p}\|j'_{\rm p}\rangle
\biggl\{\begin{array}{ccc}
j_{\rm n}&j'_{\rm n}&1\\j_{\rm p}&j'_{\rm p}&J
\end{array}\biggr\},
\label{e_me2jnp}
\end{equation}
in terms of the reduced matrix elements
\begin{equation}
\langle j\|\vec S\|j'\rangle=
-(-)^{l+1/2+j}\left[\frac{3(2j+1)(2j'+1)}{2}\right]^{1/2}
\biggl\{\begin{array}{ccc}
{\sfrac12}&{\sfrac12}&1\\j&j'&l
\end{array}\biggr\}.
\label{e_rmes}
\end{equation}
If the neutron and proton occupy the same shell, $j_{\rm n}=j_{\rm p}\equiv j$,
the matrix element reduces to
\begin{equation}
\langle j_{\rm n} j_{\rm p} J|\vec S_{\rm n}\cdot\vec S_{\rm p}|j_{\rm n} j_{\rm p} J\rangle=
\frac{J(J+1)-2j(j+1)}{2(2l+1)^2}.
\label{e_me2jnps}
\end{equation}
For the deuteron $l=0$ and $j={\sfrac12}$,
and one recovers the familiar values of $-{\frac34}$ for $J=0$ (isovector or spin singlet)
and $+{\frac14}$ for $J=1$ (isoscalar or spin triplet).

Finally, it is of use to find the reduced matrix elements in $LST$ coupling
of the separate isoscalar and isotensor parts of $\vec S_{\rm n}\cdot\vec S_{\rm p}$.
We write
\begin{equation}
\vec S_{\rm n}\cdot\vec S_{\rm p}=
T^{(000)}_{000}+T^{(002)}_{000},
\label{e_ss02}
\end{equation}
where the upper indices refer to the tensor character in $LST$
and the lower indices to the projections $M_LM_SM_T$.
The following relations are valid
\begin{align}
&\langle l^2LST=M_T=0|\vec S_{\rm n}\cdot\vec S_{\rm p}|l^2LST=M_T=0\rangle
\nonumber\\&\qquad=
\frac{\langle l^2LST=0\|\vec T^{(000)}\|l^2LST=0\rangle}{\sqrt{(2L+1)(2S+1)}},
\nonumber\\
&\langle l^2LST=1,M_T=0|\vec S_{\rm n}\cdot\vec S_{\rm p}|l^2LST=1,M_T=0\rangle
\nonumber\\&\qquad=
\frac{\langle l^2LST=1\|\vec T^{(000)}\|l^2LST=1\rangle}{\sqrt{3(2L+1)(2S+1)}}-
\frac{2\langle l^2LST=1\|\vec T^{(002)}\|l^2LST=1\rangle}{\sqrt{30(2L+1)(2S+1)}},
\nonumber\\
&\langle l^2LST=1,M_T=\pm1|\vec S_{\rm n}\cdot\vec S_{\rm p}|l^2LST=1,M_T=\pm1\rangle
\nonumber\\&\qquad=
\frac{\langle l^2LST=1\|\vec T^{(000)}\|l^2LST=1\rangle}{\sqrt{3(2L+1)(2S+1)}}+
\frac{\langle l^2LST=1\|\vec T^{(002)}\|l^2LST=1\rangle}{\sqrt{30(2L+1)(2S+1)}},
\label{e_rmelst1}
\end{align}
where the double-barred matrix elements are reduced in $L$, $S$ and $T$.
With the help of the expressions~(\ref{e_me2lst}) one deduces
\begin{align}
\langle l^2LST\|\vec T^{(000)}\|l^2LST\rangle={}&
\left[\frac{(2L+1)(2S+1)}{16(2T+1)}\right]^{1/2}[2S(S+1)-3],
\nonumber\\
\langle l^2LST\|\vec T^{(002)}\|l^2LST\rangle={}&
-\delta_{T1}\left[\frac{5(2L+1)(2S+1)}{96}\right]^{1/2}[2S(S+1)-3].
\label{e_rmelst2}
\end{align}

\section{Schematic model}
\label{s_sm}
We consider a single-$l$ shell,
corresponding to two $j$ shells, $j=l\pm{\sfrac12}$,
together with the schematic Hamiltonian
\begin{equation}
H=
\epsilon_-n_-+\epsilon_+n_+-
4\pi\sum_{T=0,1}a'_T\sum_{i<j}\delta(\vec r_i-\vec r_j)\delta(\vec r_i-R_0),
\label{e_ham}
\end{equation}
where $n_\pm$ are the number operators for the $j=l\pm{\sfrac12}$ shells
and the last term represents a surface delta interaction (SDI).
Following Brussaard and Glaudemans~\cite{Brussaard77}
we introduce the isoscalar and isovector strengths, $a_T=a'_T C(R_0)$,
where $C(R_0)$ is a radial integral,
and we adopt the notation $a\,x\equiv a_0$ and $a(1-x)\equiv a_1$,
so that $x$ indicates the relative importance of both strengths.
We note that, as long as one considers a single-$l$ or single-$j$ shell,
as is done in the following,
results obtained with SDI are identical to those with a delta interaction,
except for an overall scaling of the strengths.
We also note that the additional terms introduced in the {\em modified} SDI,
although important to reproduce nuclear binding energies~\cite{Brussaard77},
do not alter wave functions
and therefore do not influence expectation values of $\vec S_{\rm n}\cdot\vec S_{\rm p}$.
For any combination of its parameters the eigenstates of the Hamiltonian~(\ref{e_ham})
carry good angular momentum $J$ and isospin $T$.
For such eigenstates the expectation value is calculated of the operator $\vec S_{\rm n}\cdot\vec S_{\rm p}$,
which, as shown above, contains an isoscalar and an isotensor component.

The spectrum of the Hamiltonian~(\ref{e_ham}) depends on four parameters
whereas relative energies are determined
by the three parameters $\Delta\epsilon\equiv\epsilon_--\epsilon_+$, $a$ and $x$.
Eigenfunctions depend on only two dimensionless parameters $\Delta\epsilon/a$ and $x$,
which are varied in order to study their influence
on the expectation value of $\vec S_{\rm n}\cdot\vec S_{\rm p}$.
A bounded parameter can be defined as
\begin{equation}
y\equiv\frac{\Delta\epsilon/a}{5+|\Delta\epsilon/a|}.
\label{e_ypar}
\end{equation}
In most cases rapid changes in the expectation value $\langle\vec S_{\rm n}\cdot\vec S_{\rm p}\rangle$
occur for $|\Delta\epsilon/a|\approx5$ (see Sect.~\ref{ss_spinorbit}).
With the choice of 5 in the denominator this corresponds to $|y|\approx0.5$.
In the convention of a positive strength $a$ for an attractive force
and with a spin--orbit interaction that favours
the alignment of spin and orbital angular momentum ($\epsilon_-\geq\epsilon_+$),
its domain is $0\leq y\leq1$.

Calculations can be restricted to the lower half of the $l$ shell
because the results for the upper half can be obtained
through the application of a particle--hole transformation.
The Hamiltonian~(\ref{e_ham}) is not invariant under particle--hole conjugation
since this transformation induces the change $\epsilon_\pm\rightarrow-\epsilon_\pm$.
In the $(x,y)$ parametrisation introduced above
the particle--hole transformation leaves $x$ invariant and induces a sign change in $y$.
We may therefore restrict calculations to the lower half of the $l$ shell
provided we extend the parameter domain to $-1\leq y\leq+1$.
This covers all possible parameter values
for all possible nucleon numbers.

A number of limiting cases of interest occur,
which are illustrated in the next subsections.

\subsection{SU(4) symmetry}
\label{ss_su4}
If $a_0=a_1$ and $\epsilon_-=\epsilon_+$ (or $x={\frac12}$ and $y=0$),
the Hamiltonian~(\ref{e_ham}) conserves orbital angular momentum $L$, spin $S$, isospin $T$
and in addition has an SU(4) symmetry.
Since $\vec S_{\rm n}\cdot\vec S_{\rm p}$ can be written in terms of SU(4) generators,
its expectation value in the ground state depends solely
on the supermultiplet labels $(\lambda\mu\nu)$ and on $(LST)$ in the ground state.
For example, even--even $N=Z$ nuclei
have the ground-state labels $(\lambda\mu\nu)=(000)$ and $(LST)=(000)$.
Odd--odd $N=Z$ nuclei have a ground-state configuration with $(\lambda\mu\nu)=(010)$,
which contains two degenerate states
with $(LST)=(010)$ (isoscalar) or $(001)$ (isovector).
These labels completely determine the expectation value of $\vec S_{\rm n}\cdot\vec S_{\rm p}$,
which therefore is independent of the nucleon number.
For a SDI all $N=Z$ nuclei have $L=0$ in the ground state.
Denoting the ground state of an even--even $N=Z$ nucleus as $|l^{4k}L=0ST\rangle$
and that of an odd--odd $N=Z$ nucleus as $|l^{4k+2}L=0ST\rangle$,
we conclude that the following expectation values are valid:
\begin{align}
\langle l^{4k}000|\vec S_{\rm n}\cdot\vec S_{\rm p}|l^{4k}000\rangle={}&0,
\qquad
N=Z\;{\rm even},
\nonumber\\
\langle l^{4k+2}010|\vec S_{\rm n}\cdot\vec S_{\rm p}|l^{4k+2}010\rangle={}&+{\frac14},
\qquad
N=Z\;{\rm odd},
\nonumber\\
\langle l^{4k+2}001|\vec S_{\rm n}\cdot\vec S_{\rm p}|l^{4k+2}001\rangle={}&-{\frac34},
\qquad
N=Z\;{\rm odd}.
\label{e_evss0e}
\end{align}
As far as the expectation value of $\vec S_{\rm n}\cdot\vec S_{\rm p}$ is concerned,
the ground state of an odd--odd $N=Z$ nucleus
therefore behaves as a deuteron by virtue of the SU(4) symmetry.

\subsection{$LS$ coupling}
\label{ss_ls}
If $a_0\neq a_1$ and $\epsilon_-=\epsilon_+$ (or $x\neq{\frac12}$ and $y=0$),
the Hamiltonian~(\ref{e_ham}) breaks SU(4) symmetry
but conserves orbital angular momentum $L$, spin $S$ and isospin $T$.
The energy matrix associated with the Hamiltonian~(\ref{e_ham})
can therefore be constructed in an $LST$ basis.
The $\vec S_{\rm n}\cdot\vec S_{\rm p}$ operator
is not an $LST$ scalar, however,
since it has an isoscalar as well as an isotensor piece.
Its matrix elements can be calculated
from the application of the Wigner--Eckart theorem~\cite{Shalit63,Talmi93}
\begin{align}
\langle l^nLSTM_T|\vec S_{\rm n}\cdot\vec S_{\rm p}|l^nLSTM_T\rangle={}&
\frac{1}{\sqrt{(2L+1)(2S+1)(2T+1)}}
\langle l^nLST\|\vec T^{(000)}\|l^nLST\rangle
\nonumber\\
&+\frac{(-)^{T-M_T}}{\sqrt{(2L+1)(2S+1)}}
\biggl(\begin{array}{ccc}
T&2&T\\-M_T&0&M_T
\end{array}\biggr)
\langle l^nLST\|\vec T^{(002)}\|l^nLST\rangle.
\label{e_evss0u}
\end{align}
The $n$-particle $LST$-reduced matrix elements of $\vec T^{(000)}$ and $\vec T^{(002)}$
can be related recursively to the two-particle matrix elements~(\ref{e_rmelst2})
by means of coefficients of fractional parentage (CFPs) in $LST$ coupling.

The above method has the advantage of requiring
the diagonalisation of matrices of only modest dimension
but it has the drawback that CFPs have to be calculated recursively in $LST$ coupling
for the {\em total} number of nucleons.
It is therefore more efficient to consider the problem in a neutron--proton $LS$ basis.
In this basis matrices are still of reasonable dimension
and CFPs can be evaluated for the neutrons and the protons separately.
For example, for 5 (7) neutrons and 5 (7) protons in the $d$ ($f$) shell with $L=0$,
the dimensions are 26 (731) for $S=0$ and 42 (1407) for $S=1$.

\begin{figure}
\centering
\includegraphics[width=12cm]{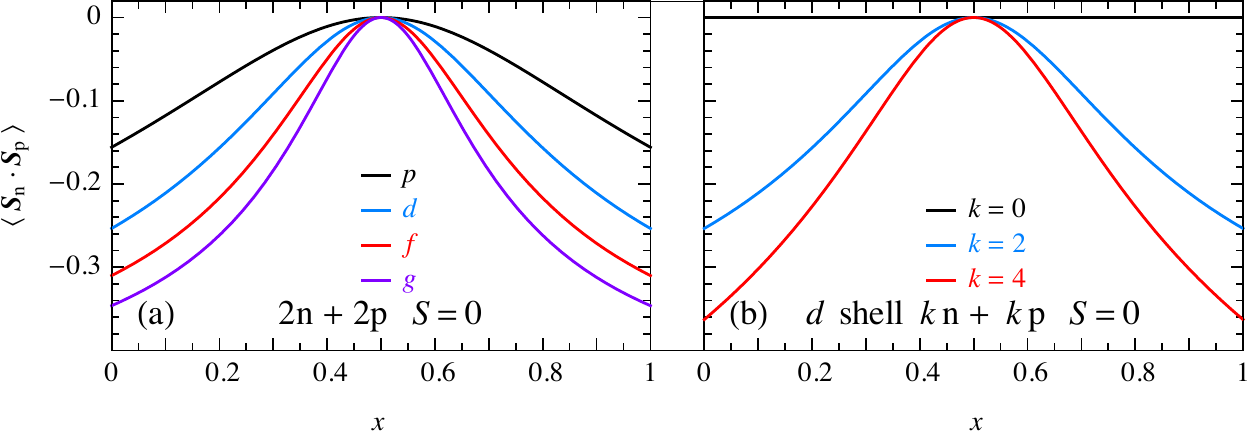}
\caption{ 
The expectation value $\langle\vec S_{\rm n}\cdot\vec S_{\rm p}\rangle$
as a function of $x=a_0/(a_0+a_1)$
in the yrast $L=0$ eigenstate of the Hamiltonian~(\ref{e_ham}) for even--even $N=Z$ systems.
(a) Ground state for two neutrons and two protons
in a $p$ (black), $d$ (blue), $f$ (red) or $g$ (purple) shell.
(b) Ground state for $k$ neutrons and $k$ protons in the $d$ shell
with $k=0$ (black), $k=2$ (blue) and $k=4$ (red).}
\label{f_leex} 
\end{figure}
Figure~\ref{f_leex} shows the expectation value $\langle\vec S_{\rm n}\cdot\vec S_{\rm p}\rangle$
as a function of $x=a_0/(a_0+a_1)$
in the ground state of $N=Z$ nuclei,
for two neutrons and two protons in a $p$, $d$, $f$ or $g$ shell
and for even numbers of neutrons and protons in the $d$ shell.
The ground state has $(LST)=(000)$ for the entire parameter range.
The expectation value $\langle\vec S_{\rm n}\cdot\vec S_{\rm p}\rangle$
is 0 at $x={\frac12}$, its value in the SU(4) limit,
and becomes more negative as the $l$ of the shell
and/or the number of nucleons increases.
Note that for $l=0$
two neutrons and two protons fill the $s$ shell (not shown in Fig.~\ref{f_leex})
and $\langle\vec S_{\rm n}\cdot\vec S_{\rm p}\rangle=0$,
independent of the Hamiltonian.
It should also be noted that $\langle\vec S_{\rm n}\cdot\vec S_{\rm p}\rangle$
is invariant under the exchange of $a_0$ and $a_1$.

\begin{figure}
\centering
\includegraphics[width=12cm]{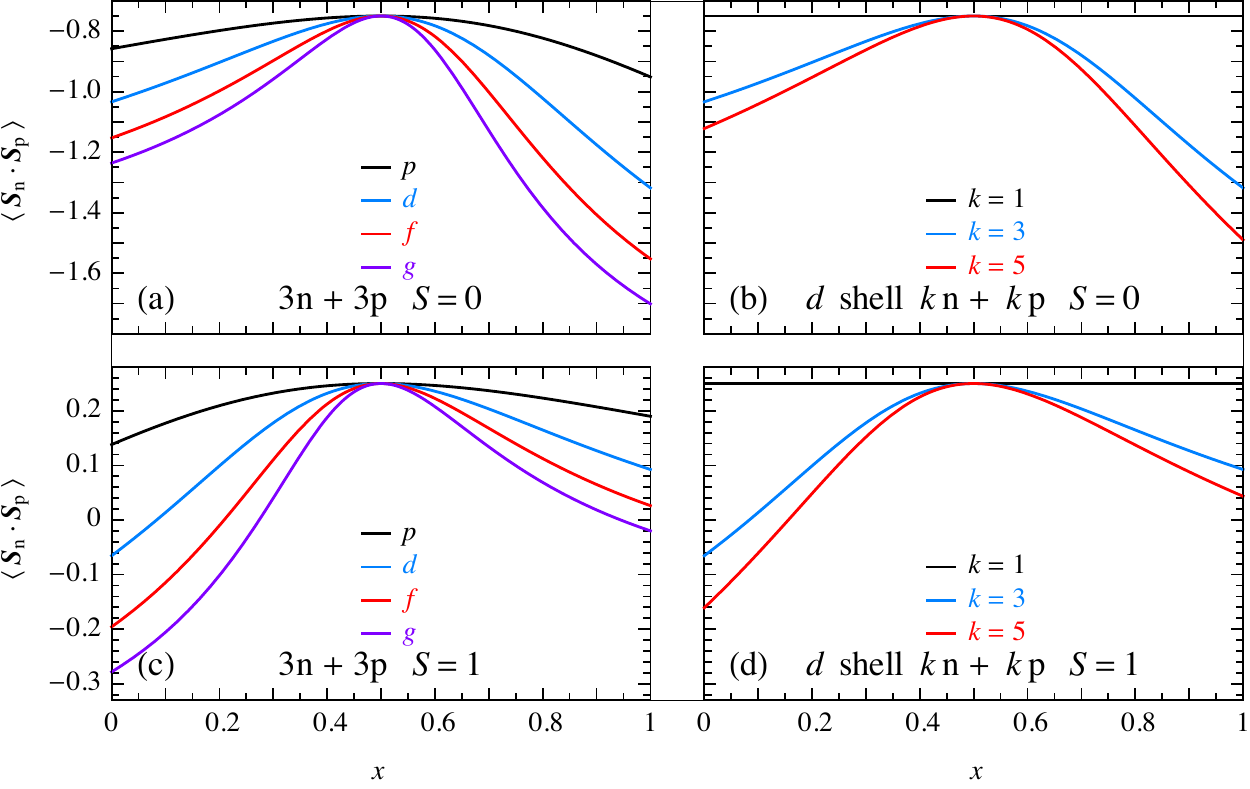}
\caption{ 
The expectation value $\langle\vec S_{\rm n}\cdot\vec S_{\rm p}\rangle$
as a function of $x=a_0/(a_0+a_1)$
in yrast $L=0$ eigenstates of the Hamiltonian~(\ref{e_ham}) for odd--odd $N=Z$ systems.
(a,c) Eigenstates with (a) $S=0$ and (c) $S=1$
for three neutrons and three protons
in a $p$ (black), $d$ (blue), $f$ (red) or $g$ (purple) shell.
(b,d) Eigenstates with (b) $S=0$ and (d) $S=1$
for $k$ neutrons and $k$ protons in the $d$ shell
with $k=1$ (black), $k=3$ (blue) and $k=5$ (red).}
\label{f_loox}
\end{figure}
Although no data are available at present for odd--odd $N=Z$ nuclei,
for completeness we show in Fig.~\ref{f_loox}
$\langle\vec S_{\rm n}\cdot\vec S_{\rm p}\rangle$
as a function of $x$ in the yrast eigenstates with $(LST)=(001)$ and $(010)$,
for three neutrons and three protons in a $p$, $d$, $f$ or $g$ shell
and for odd numbers of neutrons and protons in the $d$ shell.
For $x>{\frac12}$ the isoscalar interaction is dominant
and the ground state has $(LST)=(010)$;
for $x<{\frac12}$ the isovector interaction is dominant
and the ground state has $(LST)=(001)$.
Figure~\ref{f_loox} shows $\langle\vec S_{\rm n}\cdot\vec S_{\rm p}\rangle$
for both states over the entire range of values $0\leq x\leq1$.
For all $x$, $\langle\vec S_{\rm n}\cdot\vec S_{\rm p}\rangle$
is below its value at $x={\frac12}$,
where one recovers the SU(4) values $-{\frac34}$ and ${\frac14}$
for $S=0$ and $S=1$, respectively.
As the $l$ of the shell and/or the number of nucleons increases,
$\langle\vec S_{\rm n}\cdot\vec S_{\rm p}\rangle$ further decreases.

\subsection{Spin--orbit interaction}
\label{ss_spinorbit}
The single-particle energies of the $l+{\sfrac12}$ and $l-{\sfrac12}$ shells
are not expected to be degenerate
and, because of the spin--orbit component of the nuclear interaction,
the former is the lowest, $\epsilon_+<\epsilon_-$.
We assume for simplicity in this subsection
that the isoscalar and isovector strengths are the same, $a_0=a_1$.

\begin{figure}
\centering
\includegraphics[width=12cm]{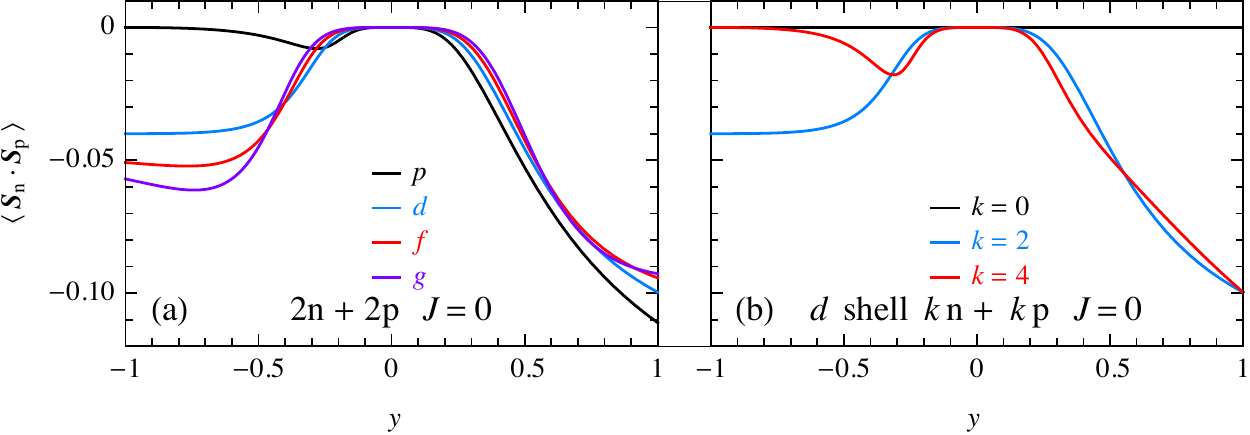}
\caption{ 
The expectation value $\langle\vec S_{\rm n}\cdot\vec S_{\rm p}\rangle$
as a function of $y$ defined in Eq.~(\ref{e_ypar})
in the $J=0$ ground state of the Hamiltonian~(\ref{e_ham}) for even--even $N=Z$ systems.
(a) Ground state for two neutrons and two protons
in a $p$ (black), $d$ (blue), $f$ (red) or $g$ (purple) shell.
(b) Ground state for $k$ neutrons and $k$ protons in the $d$ shell
with $k=0$ (black), $k=2$ (blue) and $k=4$ (red).
The isoscalar and isovector strengths are assumed equal, $a_0=a_1$.}
\label{f_leey} 
\end{figure}
Figure~\ref{f_leey} shows the expectation value $\langle\vec S_{\rm n}\cdot\vec S_{\rm p}\rangle$
in the $J=0$ ground state of the Hamiltonian~(\ref{e_ham})
for two neutrons and two protons in a $p$, $d$, $f$ or $g$ shell,
and for even numbers of neutrons and protons in the $d$ shell.
Two neutrons and two protons fill the $s$ shell (not shown in Fig.~\ref{f_leex})
and $\langle\vec S_{\rm n}\cdot\vec S_{\rm p}\rangle=0$,
independent of the Hamiltonian.
For $\epsilon_-\neq\epsilon_+$
the quantum numbers $L$ and $S$ are not conserved
and one has to revert to labelling states with their total angular momentum $J$.
Given the definition~(\ref{e_ypar}),
results for $y\rightarrow\pm1$ approach those for a single shell with $j=l\pm{\sfrac12}$.
This explains some of the values observed in Fig.~\ref{f_leey} at the limits $y=\pm1$.
For example, two neutrons and two protons fill the $p_{1/2}$ shell
and therefore the $p$ (black) curve in Fig.~\ref{f_leey}(a)
necessarily must converge to 0 at $y=-1$.
Likewise, four neutrons and four protons fill the $d_{3/2}$ shell
and the $k=4$ (red) curve in Fig.~\ref{f_leey}(b) converges to 0 at $y=-1$.
Furthermore, particle--hole symmetry explains some of the results of Fig.~\ref{f_leey}.
In a $d_{5/2}$ shell the ground state of a 2n--2p system
is the particle--hole conjugate of that of a 4n--4p system
Therefore the $k=2$ (blue) and $k=4$ (red) curves
in Fig.~\ref{f_leey}(b) converge at $y=+1$.

The spin--orbit term in the nuclear mean field is $A$ dependent
and, with use of its estimate given in Ref.~\cite{BM69},
one finds a splitting of the spin--orbit partner levels
of the order $\Delta\epsilon\approx10(2l+1)A^{-2/3}$~MeV.
The strengths of the SDI are also $A$ dependent
and a rough estimate is given in Ref.~\cite{Brussaard77}, $a_0\approx a_1\approx25A^{-1}$~MeV.
We arrive therefore at the following estimate
of the parameters of the schematic Hamiltonian~(\ref{e_ham}):
\begin{equation}
\frac{\Delta\epsilon}{a}\approx\frac{2l+1}{5}A^{1/3},
\qquad
x\approx{\frac12}.
\label{e_estimate}
\end{equation}
Application to the $d$ shell with $A^{1/3}\approx3$ leads to $|y|\approx0.375$
and we see from Fig.~\ref{f_leey} that for such values
the transition from SU(4) to the single-$j$ regime is taking place.

\begin{figure}
\centering
\includegraphics[width=12cm]{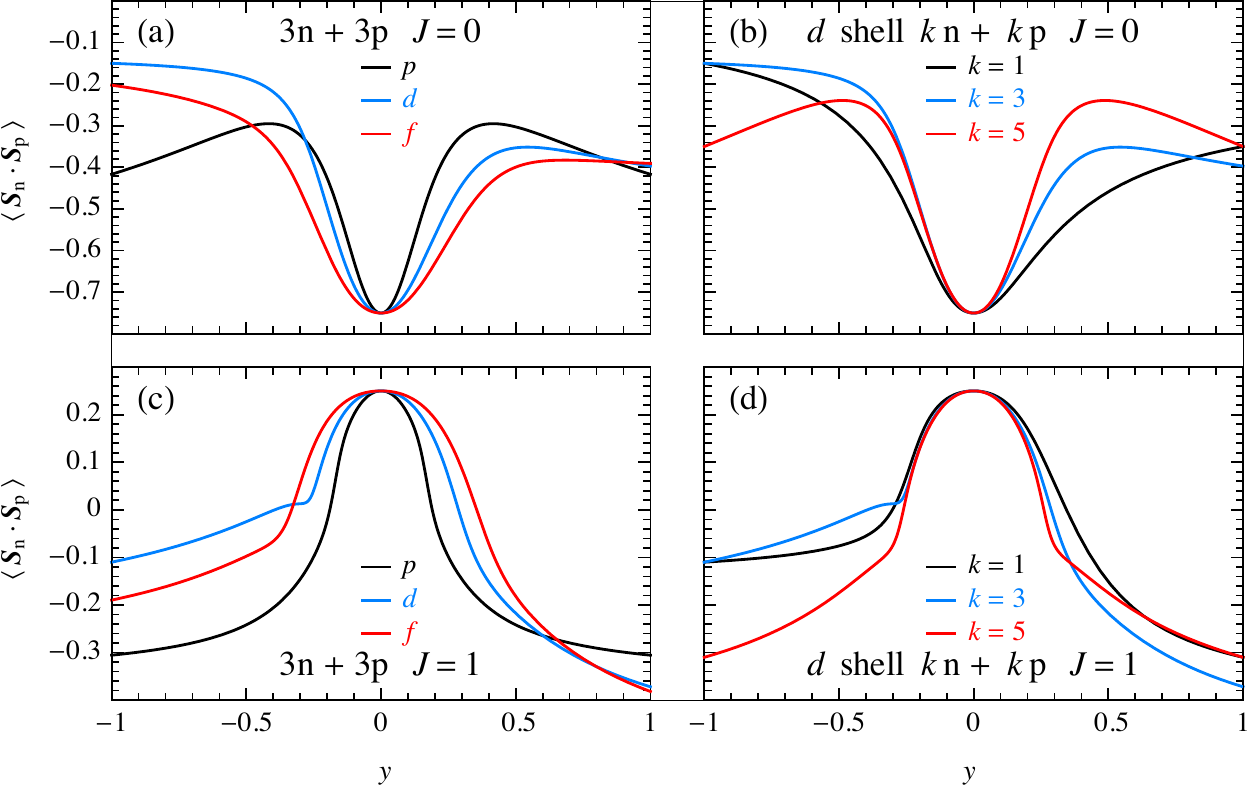}
\caption{ 
The expectation value $\langle\vec S_{\rm n}\cdot\vec S_{\rm p}\rangle$
as a function of $y$ defined in Eq.~(\ref{e_ypar})
in yrast $J=0$ and $J=1$ eigenstates of the Hamiltonian~(\ref{e_ham}) for odd--odd $N=Z$ systems.
(a,c) Eigenstates with (a) $J=0$ and (c) $J=1$
for three neutrons and three protons
in a $p$ (black), $d$ (blue) or $f$ (red) shell.
(b,d) Eigenstates with (b) $J=0$ and (d) $J=1$
for $k$ neutrons and $k$ protons in the $d$ shell
with $k=1$ (black), $k=3$ (blue) and $k=5$ (red).
The isoscalar and isovector strengths are assumed equal, $a_0=a_1$.}
\label{f_looy}
\end{figure}
For completeness we show in Fig.~\ref{f_looy} 
the expectation value $\langle\vec S_{\rm n}\cdot\vec S_{\rm p}\rangle$
as a function of $y$ in odd--odd $N=Z$ systems.
For $y=0$ one recovers the SU(4) values of $-{\frac34}$ and ${\frac14}$
for $J=0$ and $J=1$, respectively.
It is seen that systems that are self-conjugate under particle--hole symmetry,
that is, three neutrons and three protons in a $p$ shell [black curves Fig.~\ref{f_looy}(a,c)]
or five neutrons and five protons in a $d$ shell  [red curves in Fig.~\ref{f_looy}(b,d)],
display a mirror symmetry with respect to $y=0$ axis.
In the limit of infinite spin--orbit splitting, $y=\pm1$,
the model space is effectively reduced to one constructed out of a single-$j$ shell.
Particle--hole symmetry then relates $k=1$ to $k=3$ in the $d_{3/2}$ shell
[black and blue curves at $y=-1$ in Fig.~\ref{f_looy}(b,d)]
as well as $k=1$ to $k=5$ in the $d_{5/2}$ shell
[black and red curves at $y=+1$ in Fig.~\ref{f_looy}(b,d)].

\subsection{Single-$j$ shells}
\label{ss_singlej}
\begin{figure}
\centering
\includegraphics[width=12cm]{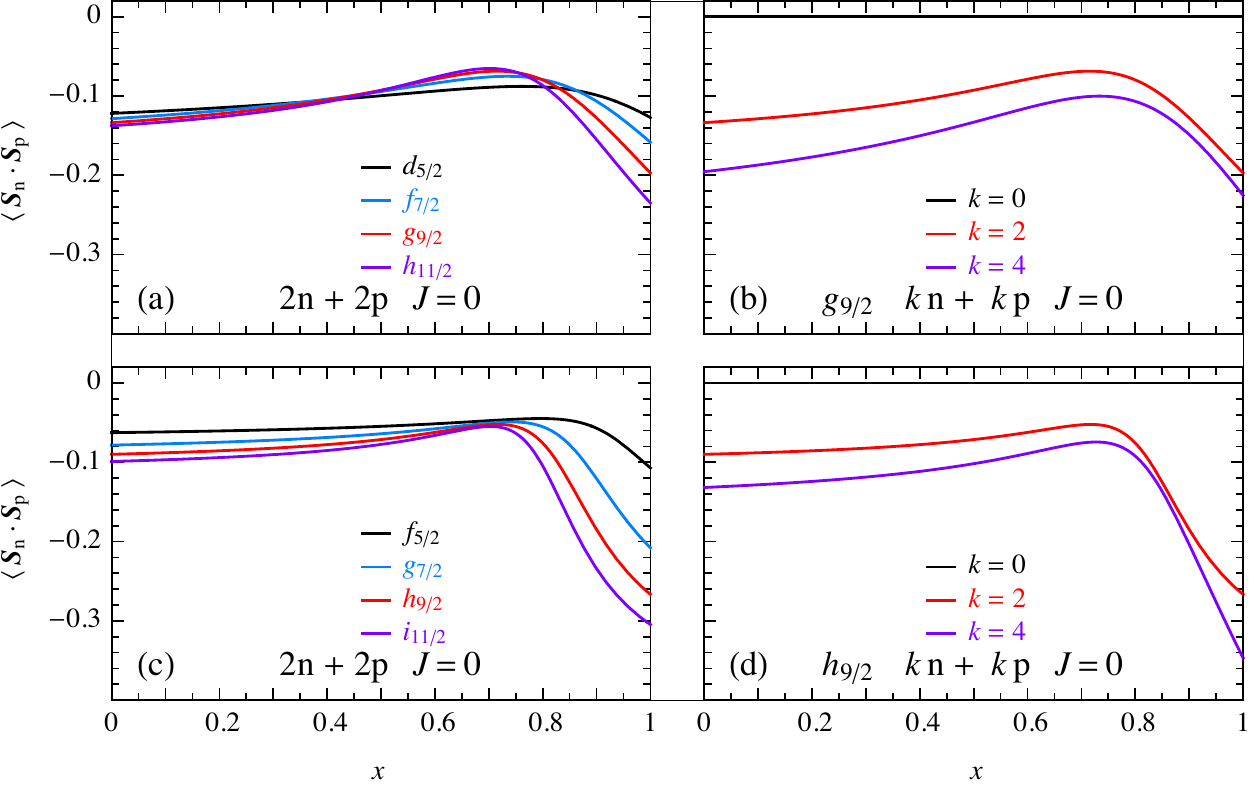}
\caption{ 
The expectation value $\langle\vec S_{\rm n}\cdot\vec S_{\rm p}\rangle$
as a function of $x=a_0/(a_0+a_1)$
in the $J=0$ ground state of the Hamiltonian~(\ref{e_ham}) for even--even $N=Z$ systems.
(a,c) Ground state for two neutrons and two protons for the
(a) $d_{5/2}$ (black), $f_{7/2}$ (blue), $g_{9/2}$ (red) and $h_{11/2}$ (purple) shell
and (c) $f_{5/2}$ (black), $g_{7/2}$ (blue), $h_{9/2}$ (red) and $i_{11/2}$ (purple) shell.
(b,d) Ground state for $k$ neutrons and $k$ protons
in the (b) $g_{9/2}$ and (d) $h_{9/2}$ shell
with $k=0$ (black), $k=2$ (red) and $k=4$ (purple).}
\label{f_jeex} 
\end{figure}
In the limit $|\epsilon_--\epsilon_+|\rightarrow+\infty$ the problem is reduced to a single-$j$ calculation.
Figure~\ref{f_jeex} shows $\langle\vec S_{\rm n}\cdot\vec S_{\rm p}\rangle$
in the $J=0$ ground state of the Hamiltonian~(\ref{e_ham})
for various even--even $N=Z$ systems confined to a single-$j$ shell.
Whether the orbital angular momentum and the spin are aligned, $j=l+{\sfrac12}$,
or anti-aligned, $j=l-{\sfrac12}$, has little influence on the results.
The expectation value is slightly less negative in the latter case
except for extreme (and unphysical) values of $x$.

\begin{figure}
\centering
\includegraphics[width=12cm]{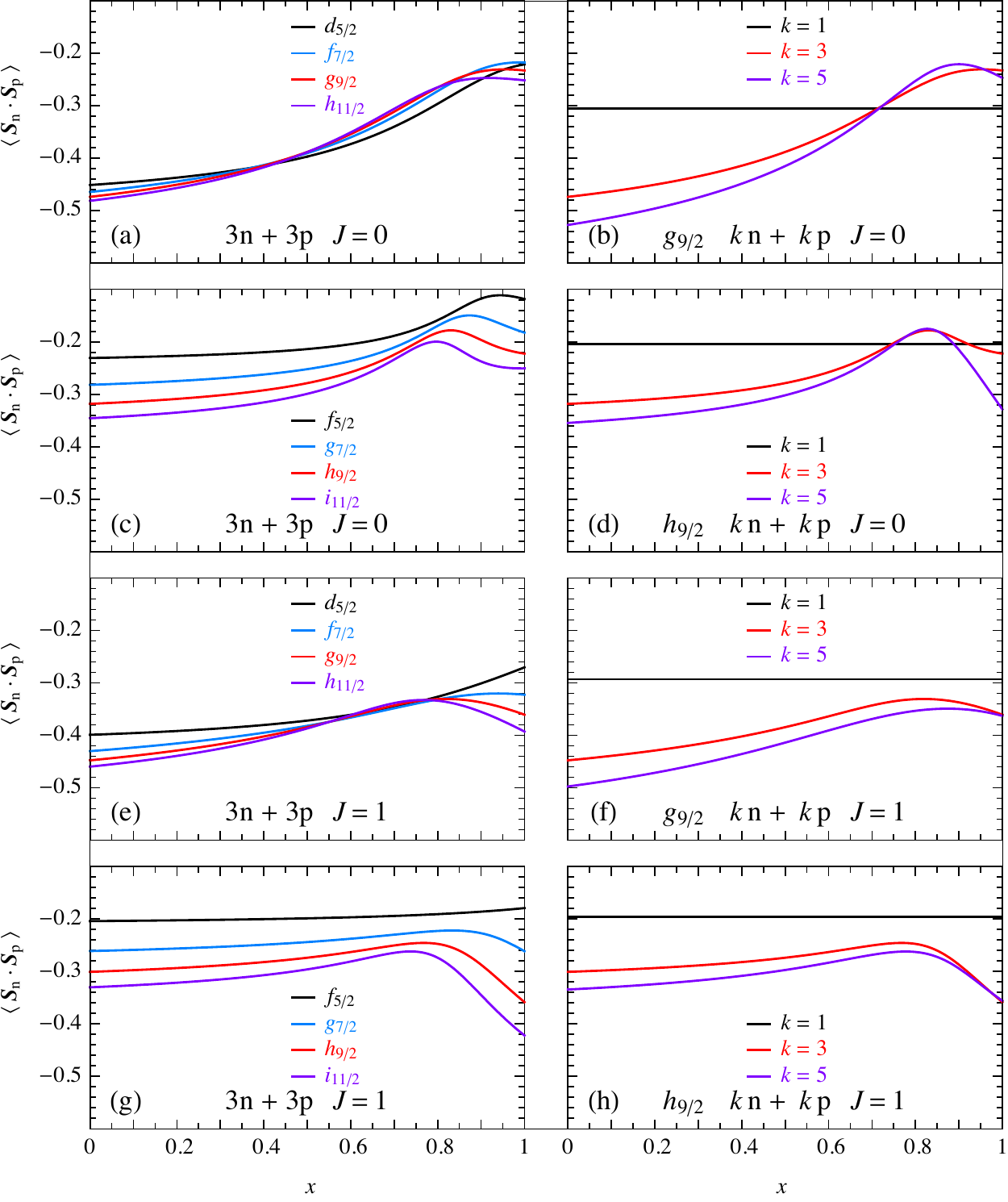}
\caption{ 
The expectation value $\langle\vec S_{\rm n}\cdot\vec S_{\rm p}\rangle$
as a function of $x=a_0/(a_0+a_1)$
in yrast $J=0$ and $J=1$ eigenstates of the Hamiltonian~(\ref{e_ham}) for odd--odd $N=Z$ systems.
(a,e) Eigenstates with (a) $J=0$ and (e) $J=1$
for three neutrons and three protons
in a $d_{5/2}$ (black), $f_{7/2}$ (blue), $g_{9/2}$ (red) or $h_{11/2}$ (purple) shell.
(c,g) Eigenstates with (c) $J=0$ and (g) $J=1$
for three neutrons and three protons
in a $f_{5/2}$ (black), $g_{7/2}$ (blue), $h_{9/2}$ (red) or $i_{11/2}$ (purple) shell.
(b,f) Eigenstates with (b) $J=0$ and (f) $J=1$
for $k$ neutrons and $k$ protons in the $g_{9/2}$ shell
with $k=0$ (black), $k=2$ (red) and $k=4$ (purple).
(d,h) Eigenstates with (d) $J=0$ and (h) $J=1$
for $k$ neutrons and $k$ protons in the $h_{9/2}$ shell
with $k=0$ (black), $k=2$ (red) and $k=4$ (purple).}
\label{f_joox}
\end{figure}
Finally, in Fig.~\ref{f_joox} the $x$-dependence of $\langle\vec S_{\rm n}\cdot\vec S_{\rm p}\rangle$
is illustrated for a variety of odd--odd $N=Z$ systems confined to a single-$j$ shell.
For $k=1$ one recovers the result~(\ref{e_me2jnps}), that is,
$-\frac{11}{36}$ and $-\frac{95}{324}$ in the $g_{9/2}$ shell,
and $-\frac{9}{44}$ and $-\frac{95}{484}$ in the $h_{9/2}$ shell,
for $J=0$ and $J=1$, respectively.
The slightly less negative values for $j=l-{\sfrac12}$ as compared to those for $j=l+{\sfrac12}$
in the general case
can therefore be traced back to the expression~(\ref{e_me2jnps}) for two nucleons.

The main conclusion from the analysis of the single-$j$-shell case
is that the expectation value $\langle\vec S_{\rm n}\cdot\vec S_{\rm p}\rangle$
in an yrast $J=0$ or $J=1$ eigenstate is found to be negative
for all possible parameter values.

It is relevant to point out the work in Ref.~\cite{Lisetskiy99},
where the isovector $M1$ transitions in odd--odd $N=Z$ nuclei
are interpreted in terms of {\em quasi-deuteron} configurations.
Simple analytical expressions for $B(M1)$ transition strengths
derived within a single-$j$ shell approximation explain well the experimental data
for both $j=l+1/2$ and $j=l-1/2$ cases for which large and small $B(M1)$s are observed.

\section{Summary and outlook}
\label{s_sum}
This study shows that there is no `simple' explanation
for the positive values of $\langle\vec S_{\rm n}\cdot\vec S_{\rm p}\rangle$
as observed in the experiments reported in Refs.~\cite{Matsubara10,Matsubara15,Tamii15}.
For all possible parameter values in the Hamiltonian~(\ref{e_ham})
the expectation value $\langle\vec S_{\rm n}\cdot\vec S_{\rm p}\rangle$
is found to be negative in the ground state of all even--even $N=Z$ nuclei.
Admittedly, the Hamiltonian~(\ref{e_ham}) is of a schematic character
and the analysis is carried out in a single-$l$ shell.
But our results show that the naive expectation
that an increase of the isoscalar (spin-triplet) interaction strength
leads to positive values of $\langle\vec S_{\rm n}\cdot\vec S_{\rm p}\rangle$
is unfounded.
Also the role of the spin--orbit term in the nuclear mean field is clearly established
as it inevitably leads to more negative $\langle\vec S_{\rm n}\cdot\vec S_{\rm p}\rangle$ values
in even--even $N=Z$ nuclei.
The interpretation of the results for odd--odd $N=Z$ nuclei is more intricate.
While no yrast $J=0$ state is found
with positive $\langle\vec S_{\rm n}\cdot\vec S_{\rm p}\rangle$,
this might occur for yrast $J=1$ eigenstates.   

The present results call for a theoretical study in similar vein
but with a more sophisticated schematic Hamiltonian.
While realistic shell-model calculations
are able to reproduce the observed spin--spin correlations~\cite{Sagawa18,Barrett13},
it would still be worthwhile to pinpoint the exact origin
of the positive $\langle\vec S_{\rm n}\cdot\vec S_{\rm p}\rangle$ values.
The positive values of $\langle\vec S_{\rm n}\cdot\vec S_{\rm p}\rangle$,
found experimentally in $sd$-shell nuclei~\cite{Matsubara10,Matsubara15,Tamii15},
might be a consequence of mixing between configurations in the $s$ and $d$ shells,
not considered in the present work.
Alternatively, they might be due
to a non-central, in particular a tensor, component of the nuclear interaction.
As the tensor interaction to some extent acts as a negative spin--orbit term,
it is yet not clear whether its effect on $\langle\vec S_{\rm n}\cdot\vec S_{\rm p}\rangle$
is adequately represented in the schematic Hamiltonian considered in this work,
although it could be partially captured in our dimensionless parameter $y$.
Finally,  the positive $\langle\vec S_{\rm n}\cdot\vec S_{\rm p}\rangle$ values
are perhaps the result of a combination of both effects,
that is, of configuration mixing and the tensor component of the nuclear interaction.
Note that the positive values seen in $^4$He and $^{12}$C,
where the $LS$ coupling scheme could be considered a good approximation,
may indeed favour the important role of the tensor force.

This study also shows the value of extending spin--spin-correlation experiments in two directions.
One is towards odd--odd $N=Z$ nuclei
where the occurrence of $J=0,T=1$ and $J=1,T=0$ states at similar energies
might give complementary information.
In this regard, measurements on $^6$Li and $^{14}$N will be of much interest. 
Above $^{40}$Ca, a program to study (p,p') scattering
with radioactive beams in inverse kinematics,
at facilities such as RIKEN~\cite{riken}, FRIB~\cite{frib} and FAIR~\cite{fair}, 
is compelling.
A second direction is to go slightly off the $N=Z$ line.
Since the $\vec S_{\rm n}\cdot\vec S_{\rm p}$ operator
is a combination of isoscalar and isotensor parts,
the measurement of its expectation value
in the $J=0$ ground state of an even--even $N=Z+2$ nucleus
as well as in its isobaric analogue state
in the neighbouring $N=Z$ odd--odd nucleus determines the separate pieces.
Along this line, an approved experiment at iThemba~\cite{aom2}
will extend the studies of Ref.~\cite{Matsubara15}
measuring the spin--spin correlations in the ground states of $^{46,48}$Ti.
For $N>Z$ targets,  a combination of (p,p') and (d,d') scattering
is required to disentangle the IS and IV components of the $M1$ operator.

\section*{Acknowledgements}
This work is based on research supported in part by the Director, Office of Science, Office of Nuclear Physics, of the U.S. Department of Energy under Contract No. DE-AC02-05CH11231 (LBNL).

\section*{Appendix: SU(4) character of $\vec S_{\rm n}\cdot\vec S_{\rm p}$}
\label{s_app}
The tensor character of $\vec S_{\rm n}\cdot\vec S_{\rm p}$ under SU(4)
can be determined by noting that the generators~(\ref{e_su4})
transform as $[211]\equiv(101)$ tensors,
where the first refers to the notation $[f'_1f'_2f'_3]\equiv[f_1-f_4,f_2-f_4,f_3-f_4]$
in terms of the Young tableau $[f_1f_2f_3f_4]$
(for a discussion of the latter, see Ref.~\cite{Hamermesh62})
and the second is the notation of supermultiplets from Ref.~\cite{Elliott81}, 
$(\lambda\mu\nu)\equiv(f_1-f_2,f_2-f_3,f_3-f_4)$.
The tensor character of any product of the SU(4) generators
must therefore belong to
\begin{align}
[211]\times[211]={}&
[0]+[211]^2+[22]+[31]+[332]+[422],
\nonumber\\
={}&(000)+(101)^2+(020)+(210)+(012)+(202).
\label{e_su4prod}
\end{align}
Furthermore, since $\vec S_{\rm n}\cdot\vec S_{\rm p}$
is a scalar in spin and a combination of a scalar and a tensor in isospin,
it can only belong to an SU(4) irreducible representation
which contains the spin--isospin combinations $(ST)=(00)$ or $(02)$.
Given the ${\rm SU}(4)\supset{\rm SU}_S(2)\otimes{\rm SU}_T(2)$ reductions
\begin{align}
[f'_1f'_2f'_3]={}&(\lambda\mu\nu)\rightarrow\textstyle{\sum_{ST}(ST)},
\nonumber\\
[0]={}&(000)\rightarrow\mathbf{(00)},
\nonumber\\
[211]={}&(101)\rightarrow(01)+(10)+(11),
\nonumber\\
[22]={}&(020)\rightarrow\mathbf{(00)}+(11)+\mathbf{(02)}+(20),
\nonumber\\
[31]={}&(210)\rightarrow(01)+(10)+(11)+(12)+(21),
\nonumber\\
[332]={}&(012)\rightarrow(01)+(10)+(11)+(12)+(21),
\nonumber\\
[422]={}&(202)\rightarrow\mathbf{(00)}+(11)^2+\mathbf{(02)}+(20)+(12)+(21)+(22),
\label{e_su4su2su2}
\end{align}
it follows that $\vec S_{\rm n}\cdot\vec S_{\rm p}$
must be a combination of $(000)$, $(020)$ and $(202)$ SU(4) tensors.

\end{document}